\documentclass[12pt,preprint]{aastex}

\usepackage{natbib}
\usepackage{amssymb}
\usepackage{graphicx}

\begin{document}
\title{Energy Spectrum of the Electrons Accelerated by a Reconnection
       Electric Field: Exponential or Power Law?}
\author{W. J. Liu, P. F. Chen, M. D. Ding, and C. Fang}
\affil{Department of Astronomy, Nanjing University,
    Nanjing 210093, China}
\email{chenpf@nju.edu.cn}
\date{}

\begin{abstract}
The direct current (DC) electric field near the reconnection region has been proposed as an effective mechanism to accelerate protons and electrons in solar flares. A power-law energy spectrum was generally claimed in the simulations of electron acceleration by the reconnection electric field. However in most of the literature, the electric and magnetic fields were chosen independently. In this paper, we perform test-particle simulations of electron acceleration in a reconnecting magnetic field, where both the electric and magnetic fields are adopted from numerical simulations of the MHD equations. It is found that the accelerated electrons present a truncated power-law energy spectrum with an exponential tail at high energies, which is analogous to the case of diffusive shock acceleration. The influences of reconnection parameters on the spectral feature are also investigated, such as the longitudinal and transverse components of the magnetic field and the size of the current sheet. It is suggested that the DC electric field alone might not be able to reproduce the observed single or double power-law distributions.
\end{abstract}


\section{INTRODUCTION}

Particle acceleration remains a mystery in solar flares, as well as in
cosmic rays from the extrasolar systems. Observationally,
the energetic particles always show a single or double power-law
spectral behavior. Three mechanisms have been proposed
\citep[see][for reviews]{mill97, asch02}, i.e., the direct current
(DC) electric field acceleration \citep[e.g.,][]{stur68}, turbulence
(or stochastic) acceleration \citep[e.g.,][]{ml98}, and shock
acceleration \citep[e.g.,][]{blan78}. Magnetic reconnection, as the
effective mechanism for magnetic energy release in solar flares, was
demonstrated to be able to provide environments favorable for all the
above-mentioned acceleration mechanisms to work \citep{chen07}.

DC electric field provides the simplest and most direct means of
accelerating particles out of a thermal plasma \citep{holm85}, which
even allows for an analytical solution of each particle trajectory under
certain assumptions \citep{bula80}. In the case that the electric and
magnetic fields induced by the accelerated particles are negligible,
test particle simulations provide a simple and
valid approach to study the particle acceleration in a
reconnection-associated electric and magnetic fields. For example,
\citet{saka92} showed that protons and electrons can be promptly
accelerated (within 1 s) up to $\sim$70 MeV and $\sim$200 MeV,
respectively. In order to compare the resulting energy spectrum of the
DC-accelerated protons with observations, \citet{mori98} conducted test
particle simulations with a hyperbolic magnetic field and a uniform
electric field. The accelerated protons present a universal power-law
spectrum, $f(E)\sim E^{-\delta}$,
with the spectral index $\delta$ being $\sim2.0-2.2$. Thereafter, a lot
of test particle simulations have been performed for electrons as well,
through either full orbit calculations \citep{zhar04,hami05} or guiding
center calculations \citep{wood05}. Almost all of these simulations
claimed to have obtained a power-law energy spectrum, with a spectral
index around 2, similar to that for protons. It is noted, however, that
the energy spectra in many of the published simulation results deviate
significantly from a power-law profile. Moreover, according to the
thick-target model, the resulting bremsstrahlung hard X-ray (HXR)
emissions in these simulations would present a power-law energy spectrum
with a spectral index around 1. However, the RHESSI observations
indicated that the HXR emissions in most solar flares possess much
softer spectra, with the spectral indices falling in the range between 3
and 7. As discussed by \citet{chen07}, several factors may be attributed
to such a big discrepancy. For example, the parameters of the
reconnecting current sheet were chosen arbitrarily due to our poor
knowledge of the physical conditions in the localized reconnection
region in solar flares. Most importantly, the electric and magnetic
fields, which should be coupled with each other through MHD equations,
were often prescribed independently. Doing so often results in the
electric and magnetic fields that are not compatible with the MHD
equations.  Some improvements were made in recent years. For example,
\citet{wood05} derived the electric field via the Ohm's law after
assuming the distributions of the magnetic field, the velocity field,
and the resistivity. A further progress was made by \citet{hami05}, who
deduced the solution of the linearized MHD equations by assuming that
the longitudinal component of the magnetic field ($B_z$) and the
velocity ($\mathbf{v}$) are small perturbations. In the real case, both
$B_z$ and $\mathbf{v}$ can be quite large, therefore, it is necessary
to obtain the electric and magnetic fields by directly solving the MHD
equations.

In order to obtain self-consistent electric and magnetic fields, in
this paper we first perform 2.5-dimensional MHD simulations of the
resistive evolution of a current sheet. When a steady state is reached,
the electric and magnetic fields are then taken out for test particle
simulations. The paper is organized as follows. In \S{2}, the numerical
method is described, including the MHD and test particle simulations.
The shape of the resulting energy spectrum is discussed in \S{3}, and a
parameter survey is conducted showing the influences of various
parameters on the spectral features in \S{4}. A short discussion is
presented in \S{5}.

\section{PROBLEM SETUP AND NUMERICAL METHOD}

\subsection{Electric and Magnetic Field Configurations}

In order to obtain self-consistent electric and magnetic fields, we
numerically solve the following 2.5-dimensional (i.e., $\partial
/\partial z=0$), time-dependent, compressible resistive MHD equations
with a multi-step implicit scheme \citep{hu89,chen00}:

\begin{equation}
{\frac {\partial \rho}{\partial t}}+{\mathbf \nabla} \cdot (\rho {\mathbf v})=0,
\end{equation}
\begin{equation}
{\frac {\partial {\mathbf v}}{\partial t}}+({\mathbf v}\cdot {\mathbf \nabla}){\mathbf v}+
{1 \over \rho}{\mathbf \nabla} P -{1 \over \rho}{\mathbf j} \times {\mathbf B}=0,
\end{equation}
\begin{equation}
{\frac {\partial \psi}{\partial t}}+{\mathbf v}\cdot {\mathbf \nabla} \psi
-\eta \Delta \psi=0,
\end{equation}
\begin{equation}
{\frac {\partial B_z}{\partial t}}+{\mathbf v}\cdot {\mathbf \nabla} B_z+B_z
        {\mathbf \nabla}\cdot {\mathbf v}-{\mathbf B}\cdot {\mathbf \nabla} v_z
        -{\mathbf \nabla}\cdot(\eta {\mathbf \nabla}B_z)=0,
\end{equation}
\begin{equation}
{\frac {\partial T}{\partial t}}+{\mathbf v} \cdot {\mathbf \nabla} T
+(\gamma-1)T{\mathbf \nabla} \cdot {\mathbf v}-{\frac {2(\gamma-1)\eta}
{\rho \beta_0}} {\mathbf j}\cdot {\mathbf j}-CQ/\rho=0,\label{ener}
\end{equation}

\noindent where $\gamma=5/3$ is the ratio of specific heats, $\eta$ is
the dimensionless resistivity, that is, the inverse of the magnetic
Reynolds number. The last term on the left-hand side of equation
(\ref{ener}) is the field-aligned heat conduction, where $Q=\nabla \cdot
[ T^{5/2} (\mathbf{B}\cdot \nabla T)\mathbf{B}/B^2 ]$, and $C$ is the
dimensionless coefficient \citep[see][for details]{chen99}.  The five
independent variables are the density ($\rho$), velocity ($\mathbf{v}$),
magnetic flux function ($\psi$), the longitudinal component of the
magnetic field ($B_z$), and temperature ($T$); note that the magnetic
field $\mathbf{B}$ is related to the magnetic flux function through
${\mathbf B} = \nabla \times (\psi \hat{\mathbf e_z}) + B_z
\hat{\mathbf e_z}$. When nondimensionalizing the MHD equations,
the characteristic density and temperature are fixed to be
$\rho_0=1.67\times 10^{-11}$ kg m$^{-3}$ and $T_0=10^6$ K,
respectively. While, the length scale ($L_0$) and the plasma
beta ($\beta_0$) are free parameters. The velocity is normalized by
the isothermal sound speed $v_0$, and the time is therefore by $\tau_0=
L_0/v_0$. Another time scale, the Alfv\'en transit time $\tau_{\rm A}$,
is related to $\tau_0$ by $\tau_{\rm A} =\sqrt{\beta_0/2}\tau_0$.
The dimensionless resistivity $\eta$ is distributed as

\begin{equation}
\eta =\cases{\eta_0\cos (5\pi x)\cos [2(y-4)\pi], & if $|x| \leq 0.1$,
                        $|y-4| \leq 0.25$, \cr
                0, &  elsewhere,\cr}
\end{equation}
\noindent
where $\eta_0$ is also a free parameter.

As the initial conditions, uniform density ($\rho=1$) and temperature
($T=1$) are assumed. The initial magnetic field is chosen to be a
force-free field with a vertical current sheet located along the
$y$-axis, which is written as

\begin{equation}
\psi = \left\{
\begin{array}{ll}
(2 {\it w} / \pi) \cos(\pi x / 2 {\it w}) - {\it w} & (|x| \le {\it w}), \\
-|x| & ( |x| > {\it w} ),
\end{array}
\right.
\end{equation}

\begin{equation}
B_z = \left\{
\begin{array}{ll}
\sqrt{B_{g}^2 + \cos^2( \pi x / 2 {\it w} )} & (|x| \le {\it w}), \\
B_{g} & (|x| > {\it w}),
\end{array}
\right.
\end{equation}

\noindent
where the half-width of the current sheet is $w=0.1$, $B_{g}$ is the
longitudinal magnetic field (i.e., the guide component) in the
background.

The dimensionless size of the simulation box is $-3\leq x \leq 3$ and
$0\leq y \leq 8$. Because of the symmetry, the calculation is made only
in the top right quadrant. The top ($y=8$) and the right-hand ($x=3$)
sides are treated as open boundaries. Symmetry conditions are applied
to the left-hand boundary ($x=0$) and the bottom ($y=4$) of the
simulation quadrant. The detailed description of the MHD simulation can
be found in \citet{chen99}.

\begin{figure}
\begin{center}
\plotone{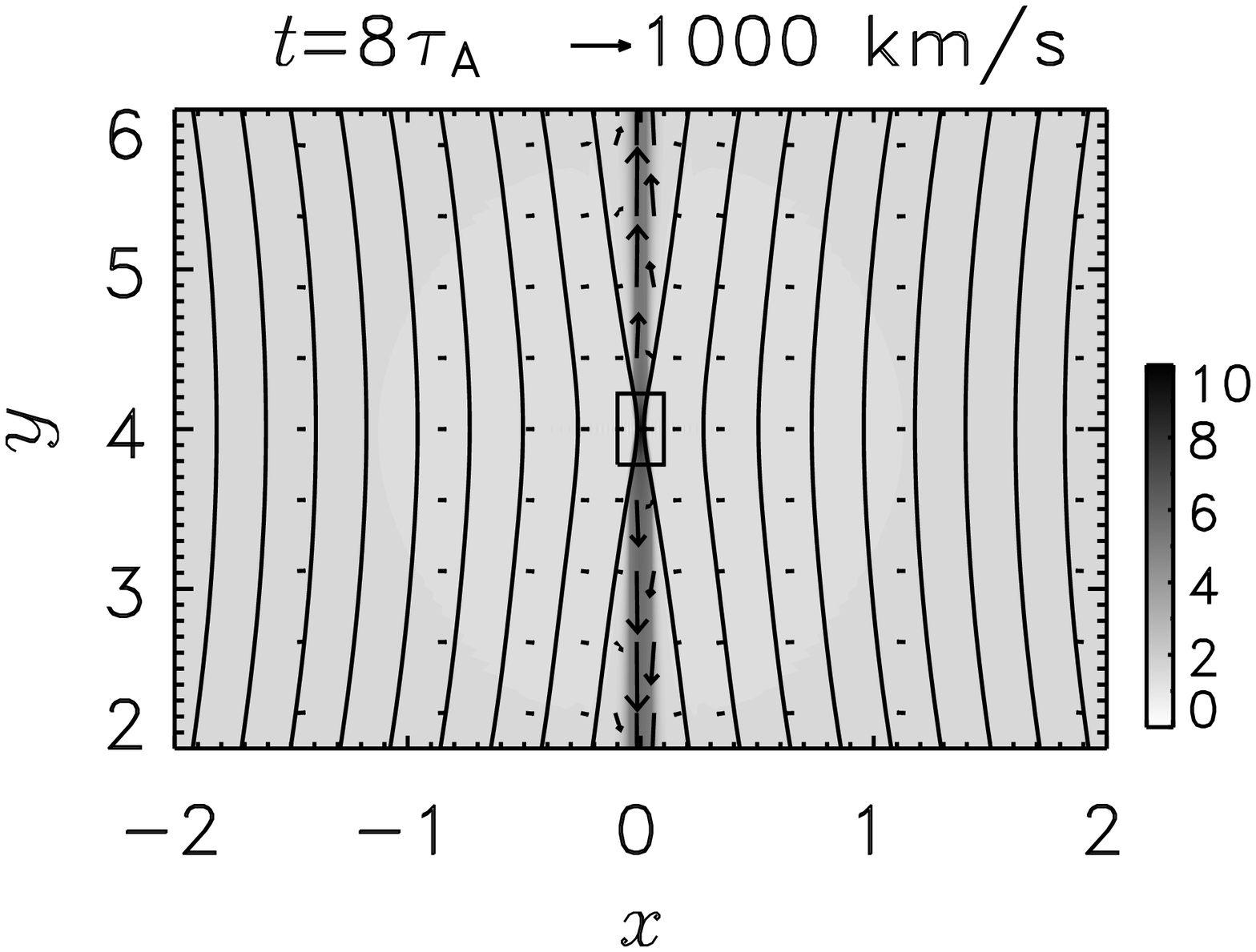}
\end{center}
\caption{Distributions of the temperature ({\it gray scale}), the
magnetic field ({\it solid lines}), and the velocity ({\it vector
arrows}) at $t=8\tau_{\rm A}$. The rectangle around the reconnection
X-point is the site where test particles are injected.}\label{f1}
\end{figure}

As the localized resistivity sets in, the elongated current sheet
dissipates and collapses into an X-type magnetic configuration. After
$\sim 7\tau_{\rm A}$, the dynamics of the whole simulation region
becomes steady, keeping all physical quantities almost invariant with
time. Figure \ref{f1} shows the distributions of the temperature ({\it
gray scale}), the magnetic field ({\it solid lines}), and the velocity
({\it vector arrows}) for the case with $B_{g}=1$, $\beta_0=0.01$, and
$\eta_0=0.02$ at $t=8\tau_{\rm A}$. Note that at the steady state, the
$z$-component of the magnetic field near the reconnection X-point is
around $B_{g}$. The small rectangular box in Figure \ref{f1} indicates
the region where the resistivity does not vanish and where test
particles are injected.

The corresponding electric field ($\mathbf E$) is then determined by the
Ohm's law, i.e., ${\mathbf E}=\eta \nabla \times {\mathbf B}-{\mathbf v}
\times {\mathbf B}$, where ${\mathbf E}$ is nondimensionalized by the
characteristic value of $v_0B_0$, $B_0$ is the normalization unit of the
magnetic field \citep[see][for details]{chen99}. Thereby, the electric
and magnetic fields are obtained for the ensuing test particle
simulations. As an example, in the case of $\beta=0.01$ and $\eta_0$=
0.01, the electric field at the reconnection site is $\sim 600$ 
V m$^{-1}$, which is super-Dreicer.

\subsection{Test Particle Approach}

Test electrons are uniformly distributed in the resistive region, which
is indicated by the rectangle in Figure \ref{f1}. Initially, these
electrons have a Maxwellian velocity distribution with the local plasma
temperature, which is superimposed on the local plasma bulk velocity.
The motion of each electron is then calculated by numerically solving the
following relativistic Lorentz equations

\begin{equation}
\frac{d}{d t}(\gamma m_0{\mathbf v})=q({\mathbf E}+{\mathbf v}\times
{\mathbf B} ),\label{momen}
\end{equation}

\begin{equation}
\frac{d {\mathbf x}}{d t} = {\mathbf v},
\end{equation}
\noindent
where ${\mathbf x}$ and ${\mathbf v}$ are the particle position and
velocity vectors, respectively; $\gamma =1/\sqrt{1-v^2/c^2}$ is the
Lorentz factor, $m_0$ and $q$ are the rest mass and the charge of the
electron, respectively; ${\mathbf E}$ and ${\mathbf B}$ are the electric
and magnetic fields obtained from the MHD simulations as described in
the previous subsection. It is noted here that collisions are neglected
in the momentum equation (\ref{momen}). 

The 4th-order Runge-Kutta-Fehlberg (RKF45) scheme is used to solve the
above equations, where the time step ($\Delta t$) is adaptive. In order
to get smooth energy spectra,  $2 \times 10^5$ test particles are
simulated.

In order to confirm our test particle code, we repeated the simulations
of proton acceleration that were performed by \citet{mori98}, with the
same electric and magnetic fields, i.e., ${\mathbf E}=(0, 0,
E_z)$ and ${\mathbf B} = (\alpha y,\beta x, B_z)$. We chose the same
parameters as in \citet{mori98} and found that the spectral indices are
about 1.9-2.3, almost identical to their result, i.e., 2.0-2.2.

\section{ENERGY SPECTRUM: POWER LAW OR EXPONENTIAL?}

\begin{figure}
\begin{center}
\plotone{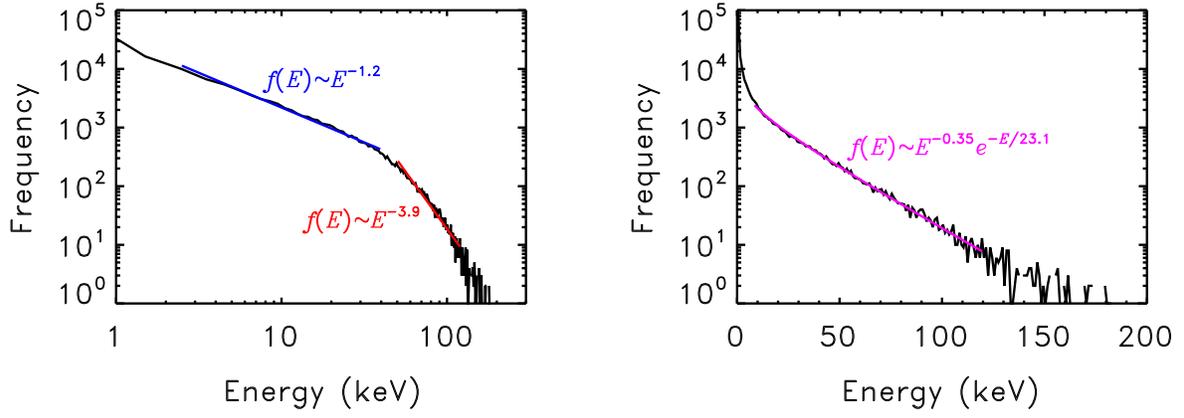}
\end{center}
\caption{Energy spectrum of the nonthermal electrons for the case with
$B_z=1$, $L_0=100$ m, $\beta_0=0.01$, and $\eta_0=0.02$ plotted in a
log-log scale ({\it left}, which is fitted with a double power law) and
a linear-log scale ({\it right}, which is fitted with a power law
truncated by an exponential tail).}\label{f2}
\end{figure}

The electrons are accelerated under the action of the DC electric
field promptly. At the same time, they gyro-rotate around the magnetic
field lines. After certain rounds of gyro-rotation, the accelerated
electrons drift out of the resistive region, and then propagate away
along magnetic separatrix layers between the reconnection inflow and
outflow. Since the acceleration and the propagation of the electrons
are 2-fold rotational symmetric, the particles going upward and
downward are collected together to construct their energy spectrum.
The left panel of Figure \ref{f2} plots the energy spectrum in a log-log
scale for the case with $B_{g}=1$, $L_0$=100 m, $\beta_0=0.01$, and
$\eta_0=0.02$. To our surprise, the spectrum, which is quite smooth
except for the high energy tail, does not show a single power-law
profile (note that a single power-law profile is manifested as a linear
line in the log-log scale). Apparently, it looks like a double power-law
profile, as proposed in our recent review paper \citep{chen07}. The
spectral profile is then fitted with a double power-law distribution,
i.e., a harder power-law spectrum $f(E)\sim E^{-1.2}$ in the energy
range of 2-40 keV ({\it red line}) plus a softer power-law spectrum
$f(E)\sim E^{-3.9}$ in the energy range of 50-120 keV ({\it blue line}).
It is noted that the spectral index at the high energy tail in our
simulation is significantly higher than those obtained in the published
test particle simulations, and is in the typical range of the electron
spectral indices derived from HXR spectral observations.

The HXR observations from RHESSI did find double (or broken) power-law
spectral profiles in some flares \citep[e.g.,][]{holm03}, and
theoretical models have also been proposed to explain this feature
\citep[e.g.,][]{zhan04}. The profile in the left panel of Figure
\ref{f2} is, however, somewhat different from the broken power-law
distribution of \citet{holm03} in that the spectral profile changes
gradually between the two power-law parts in our simulation results,
while the transition is abrupt in the observation \citep{sui07}.

In order to clarify the spectral distribution of the accelerated
electrons in our simulations, we re-plot the energy spectrum in a
linear-log scale in the right panel of Figure \ref{f2}. It is seen
that, except for the significantly enhanced lower energy tail, the
spectral profile is almost a linear line in a wide energy range,
which means that the spectrum is close to be exponential in most of
the energy range. Considering the low energy tail, we fit the
spectral profile with the combination of an exponential and a
power-law functions, i.e., $f(E)\sim E^{-\delta}e^{-E/E_0}$. With the
spectral index $\delta=0.35$ and the rollover energy $E_0$=23.1 keV,
the fitted line ({\it pink}), as shown in the right panel of Figure
\ref{f2}, is consistent with the simulation result in most of the
energy range, from 8 to over 120 keV.

Although it was claimed by many authors that DC electric field in
magnetic reconnection leads to a power-law energy spectrum of the
accelerated electrons, quite often the spectrum obtained from
test particle simulations deviates significantly from a perfect
power-law distribution. Several authors already noted that the energy
spectrum of the DC-accelerated electrons may be characterized as either
power-law or exponential, depending on the value of $B_z$ \citep{hami05}
or the trapping time of the electrons inside the acceleration volume
\citep{anas97}.  The test particle simulations in this paper, with
self-consistent electric and magnetic fields obtained from MHD
simulations, indicate that the electrons accelerated by the DC electric
field of magnetic reconnection present a spectrum with the form

\begin{equation}
f(E)\sim E^{-\delta} e^{-E/E_0},\label{eqa}
\end{equation}

\noindent
which is similar to those obtained in the diffusive shock acceleration
when the power-law behavior is truncated by a variety of effects
\citep[see][for discussions]{elli85}.  The same spectral profile was
also obtained in the electron acceleration by random DC electric field
\citep{anas02}. It is, however, mentioned here that among all our
simulated scenarios, only in one case with $B_{g}=1.0$, $L_0$=50 m,
$\beta_0=0.01$, and $\eta_0=0.005$, the energy spectrum does show a
clear broken power-law shape.

\section{PARAMETER SURVEY}

Several parameters can affect the energy spectral profile
of the DC-accelerated electrons in our simulations, where the
spectral profile is characterized by $\delta$ and $E_0$ as
indicated by equation (\ref{eqa}). The free parameters include the
longitudinal component of the magnetic field ($B_{g}$), the
length scale ($L_0$), the resistivity ($\eta_0$), and the magnetic
field strength, as represented by the plasma beta ($\beta_0$). In the
following subsections, the effect of varying each parameter is
investigated individually with other parameters keeping fixed.

\subsection{Effect of the Guide Field $B_{g}$}

\begin{figure}
\epsscale{0.7}
\plotone{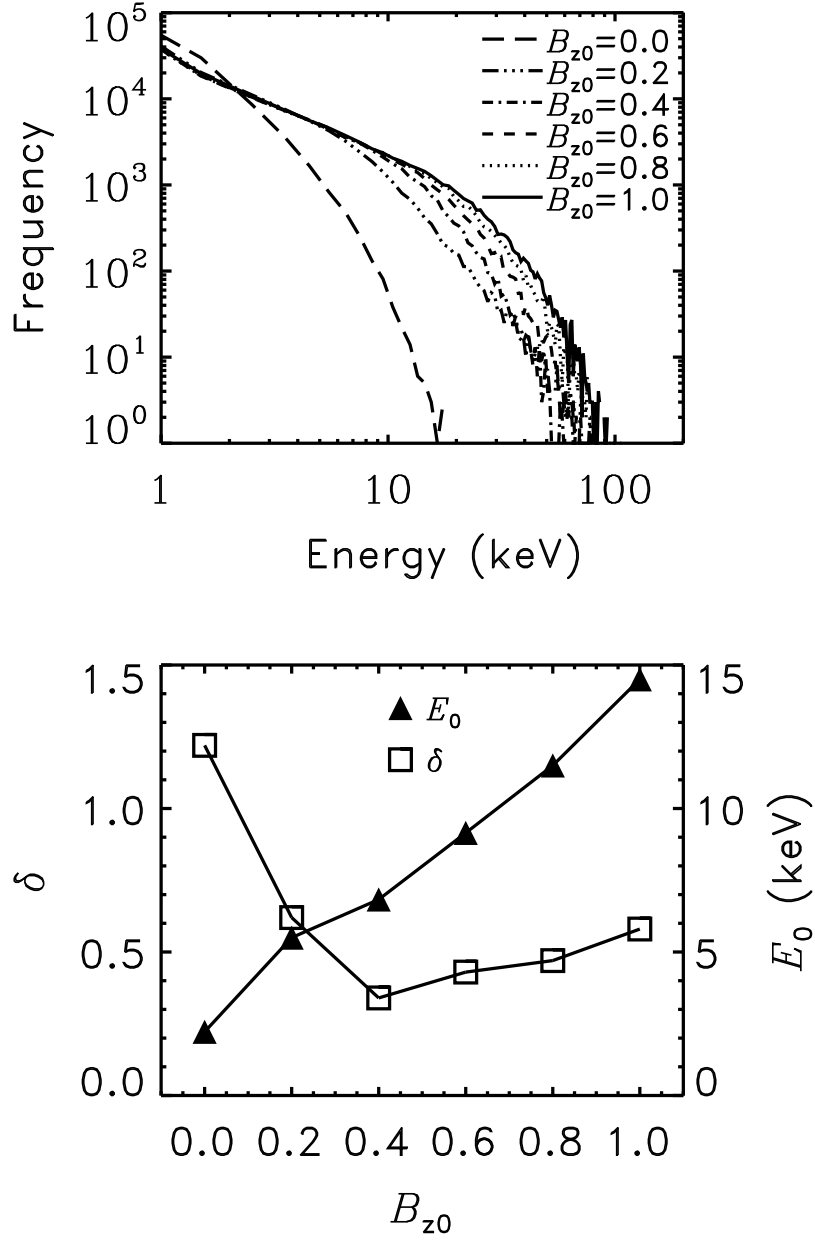}
\caption{{\it Upper}: Energy spectra of the nonthermal electrons
in six cases with different guide magnetic field $B_{g}$, where
other parameters are fixed ($L_0$=50 m, $\beta_0=0.01$, and $\eta_0=
0.02$). {\it Lower}: Variations of $\delta$ and $E_0$ with
$B_{g}$. }\label{f3}
\end{figure}

By theoretical analysis of electron acceleration in a reconnecting
current sheet, \citet{lit96} propounded that a longitudinal component
of magnetic field is necessary to explain the accelerated electrons
with energy up to 100 keV. In our simulations, the longitudinal
component (i.e., the guide component along the $z$-direction)
of the magnetic field near the reconnection X-point is around $B_{g}$.
Therefore, we calculated six cases with $B_{g}$ increasing from 0 to 1.0,
while the other parameters are $L_0$=50 m, $\beta_0=0.01$, and $\eta_0=0.02$.
The energy spectra of the accelerated electrons in the six cases
are depicted in the upper panel of Figure \ref{f3}. It is clear that,
as $B_{g}$ increases, more and more electrons are accelerated to
high energies, which confirms the importance of longitudinal magnetic
field in accelerating electrons to higher energies \citep{lit96}.
The dependence of $\delta$ and $E_0$ on $B_{g}$ is plotted in the lower
panel of Figure \ref{f3}. It is found that as $B_{g}$ increases,
$E_0$ increases steadily, while $\delta$ first falls rapidly when $B_{g}$
increases from 0 to 0.4, and then increases slowly.

In order to study the efficiency of DC acceleration, we calculate the
percentage of electrons whose energy is higher than 10 keV, and define
it as the acceleration rate. We find that when $B_{g}$ is zero, few
particles are accelerated and the acceleration rate is 0.0475\%.
However, when $B_{g}$ reaches 1, the acceleration rate jumps to 11.3\%.

\subsection{Effect of the Length Scale $L_0$}

\begin{figure}
\epsscale{0.7}
\plotone{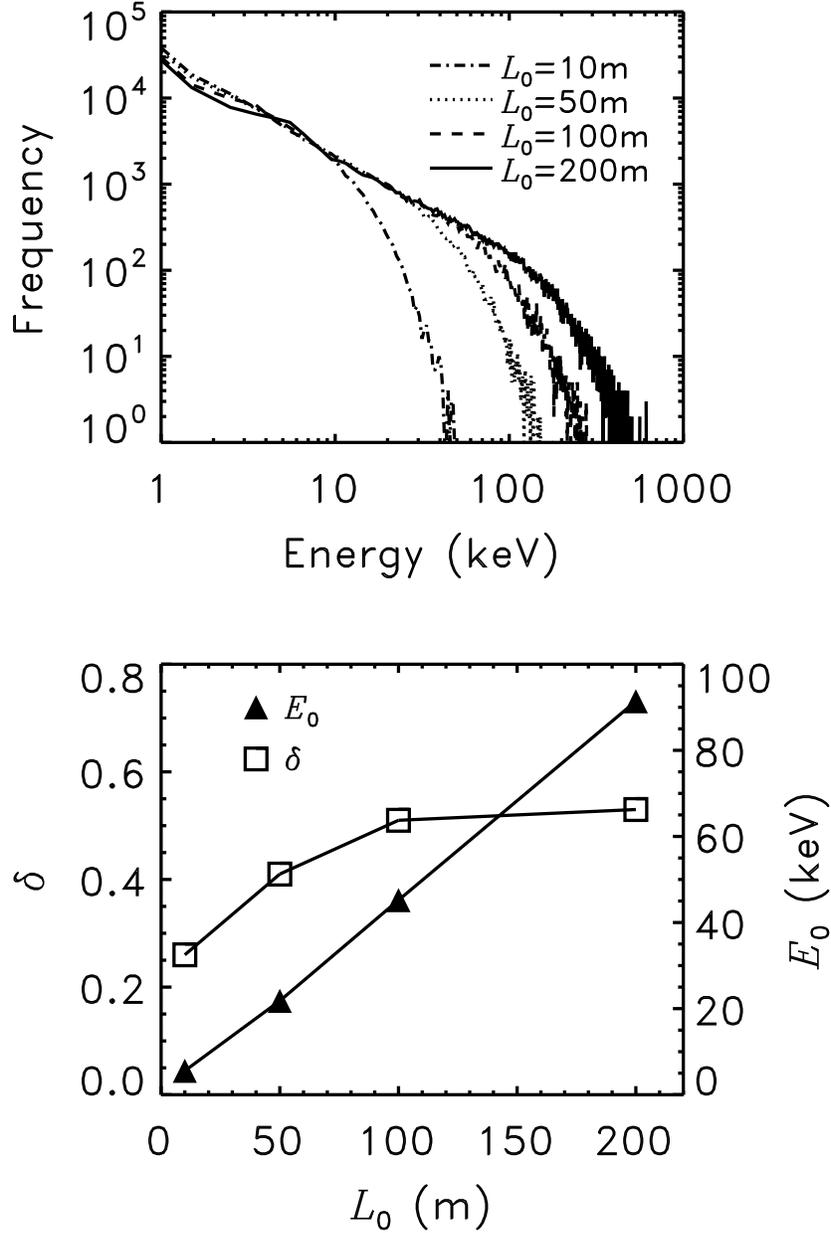}
\caption{{\it Upper}: Energy spectra of the nonthermal electrons
in four cases with different length scale $L_0$, where other parameters
are fixed ($B_{g}=1$, $\beta_0=0.005$, and $\eta_0=0.02$). {\it Lower}:
Variations of $\delta$ and $E_0$ with $L_0$.}\label{f4}
\end{figure}

The size of the current sheet is an important but unknown parameter in
the flaring process. Therefore, it is worth studying how this parameter
would affect the energy spectrum of the electrons. As shown in Figure
\ref{f1}, the initial size of the reconnection sheet is $0.2L_0\times
0.5L_0$, where $L_0$ is the length scale of the MHD simulations. With
$B_g=1$, $\beta_0=0.005$, and $\eta_0=0.02$ being fixed, test particle
simulations are conducted in four cases with $L_0$ being 10, 50,
100, and 200 m, respectively. The upper panel of Figure \ref{f4} shows
the energy spectra of the accelerated electrons in the four cases. It is
seen that as $L_0$ increases, more and more electrons are accelerated to
high energies. This is easy to understand since as $L_0$ increases,
electrons travel a longer distance in the electric field before they
migrate out of the acceleration region due to gyration. The dependencies
of $\delta$ and $E_0$ on $L_0$ are shown in the lower panel of Figure
\ref{f4}. It is seen that $E_0$ increases almost linearly with $L_0$,
while $\delta$ initially increases rapidly with $L_0$, and then
saturates after $L_0$=100 m. We calculated the fraction of the electrons
with energies exceeding 10 keV, and found that the acceleration rate
increases from 4.49\% to 26.3\% as $L_0$ increases from 10 to 200 m.

\subsection{Effect of the Plasma beta $\beta_0$}

\begin{figure}
\epsscale{0.7}
\plotone{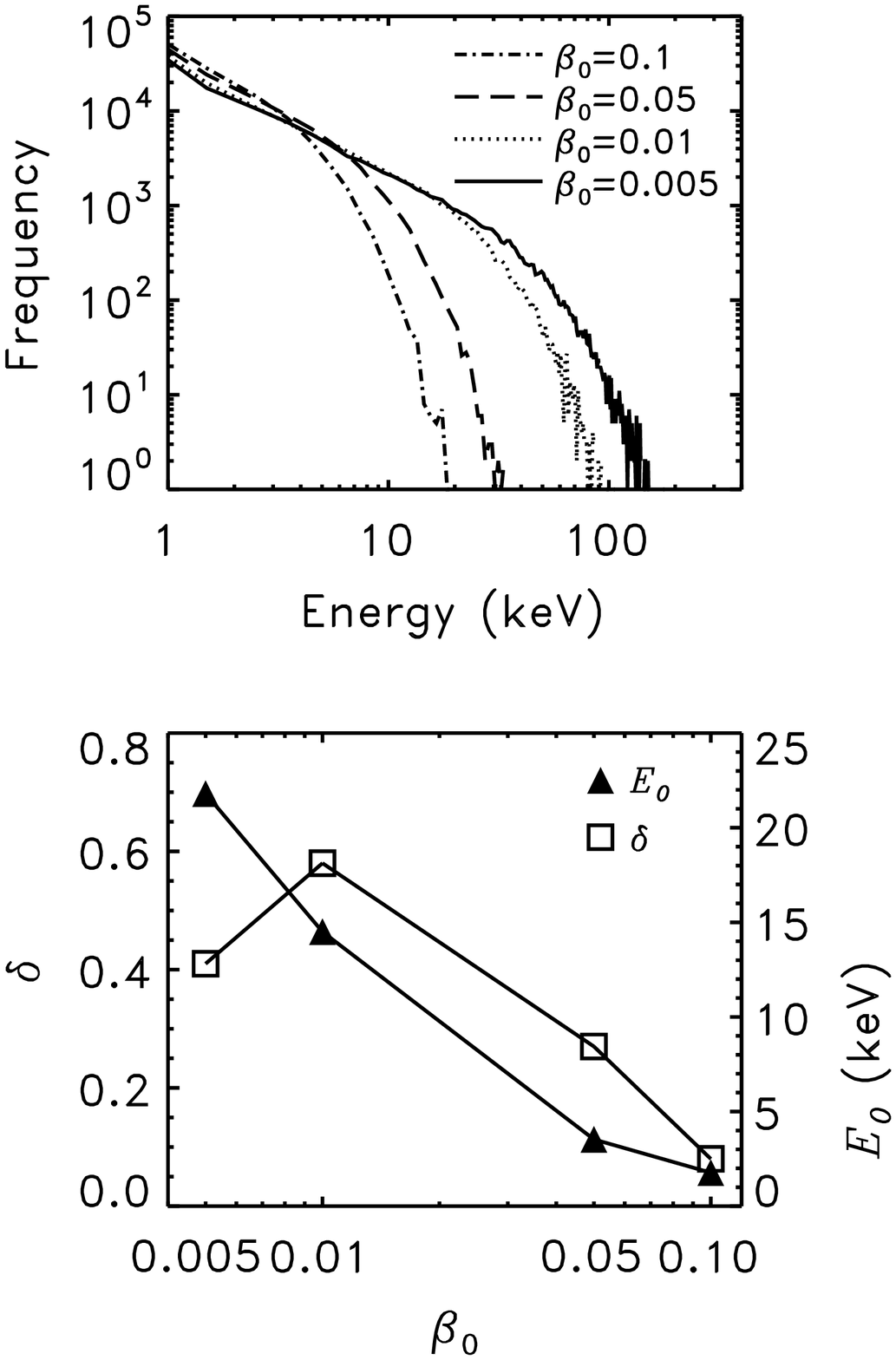}
\caption{{\it Upper}: Energy spectra of the nonthermal electrons
in four cases with different $\beta_0$, where other parameters are fixed
($B_{g}=1$, $L_0=50$ m, and $\eta_0=0.02$).  {\it Lower}: Variations
of $\delta$ and $E_0$ with $\beta_0$.}\label{f5}
\end{figure}

In the Petschek model for magnetic reconnection \citep{pets64}, the
electric field near the reconnection X-point is proportional to $B_0^2$,
where $B_0$ is the magnetic field strength in the inflow region. 
Therefore, the spectral shape of the accelerated electrons should be
sensitive to the magnetic field. Here, the magnetic field ($B_0$) is
characterized by the plasma beta, i.e., $\beta_0=2\mu_0\rho_0RT_0/B
_0^2$, With the choice of $\rho_0$ and $T_0$ as in \S 2, $B_0$ is
related to $\beta_0$ by $B_0\sim 7/\sqrt{\beta_0}$ in units of gauss.
With $B_{g}=1$, $L_0$=50 m, and $\eta_0=0.02$, we simulated four cases
with different $\beta_0$.  The upper panel of Figure \ref{f5} shows the
energy spectra of the accelerated electrons in the four cases. As
expected, it is seen that as $\beta_0$ decreases, i.e., the magnetic
field increases, more electrons are accelerated to high energies, and
the spectrum at the high energy tail tends to become harder.
The dependence of $\delta$ and $E_0$ on $\beta_0$ is plotted in the
lower panel of Figure \ref{f5}, which shows that as $\beta_0$ decreases,
$E_0$ increases significantly, while $\delta$ increases when $\beta_0$
decreases from 0.1 to 0.01 and then decreases. The acceleration rate
increases significantly from 0.169\% to 15.2\% as $\beta_0$ decreases
from 0.1 to 0.005.

\subsection{Effect of the Resistivity $\eta_0$}

\begin{figure}
\epsscale{0.7}
\plotone{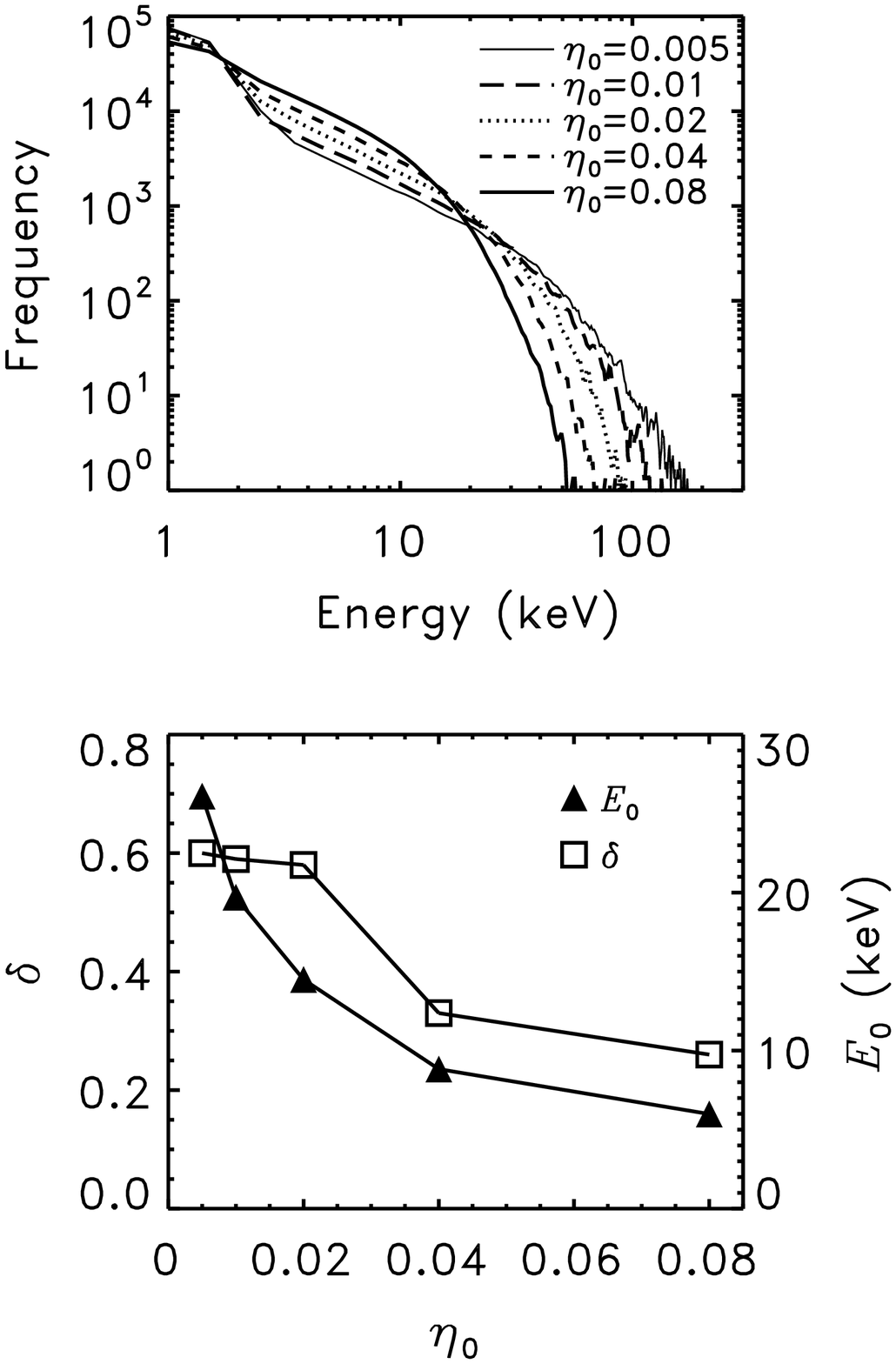}
\caption{{\it Upper}: Energy spectra of the nonthermal electrons 
in four cases with different $\eta_0$, where other parameters are fixed
($B_{g}=1$, $L_0=50$ m, and $\beta_0=0.01$).  {\it Lower}: Variations
of $\delta$ and $E_0$ with $\eta_0$.}\label{f6}
\end{figure}

The resistivity in a reconnecting current sheet is caused by the local
microscopic instability of plasma and it is still an unknown parameter
in the flaring process. In our simulations the resistivity is
characterized by the free parameter $\eta_0$. With $B_z=1$, $\beta_0=
0.01$, and $L_0=50$ m, five cases with different $\eta_0$ are simulated.
Note that $\eta_0=0.01$ corresponds to an anomalous magnetic
diffusivity of $7\times 10^4$ $\Omega$ m, $\sim 10^5$ times the classic
value, which is quite similar to the numerical results of
particle-in-cell simulations \citep{petk03,karl08}. Such an anomalous
magnetic diffusivity, in addition to the small normalization length
scale, makes the magnetic Reynolds number, i.e., the inverse of
$\eta_0$, be $\sim100$. The upper panel of Figure \ref{f6} shows the
corresponding energy spectra of the accelerated electrons. It is found
that the acceleration rate slightly increases from 10.1\% to 11.3\% as
$\eta_0$ increases from 0.005 to 0.02 and then slightly decreases to
9.70\% as $\eta_0$ increases to 0.08. The weak dependence of the
acceleration rate is actually due to the population decrease of the
electrons above $\sim 20$ keV and the population increase of the
electrons between $\sim 10$ keV and $\sim 20$ keV as $\eta_0$ increases.
The dependence of $\delta$ and $E_0$ on $\eta_0$ is depicted in the
lower panel of Figure \ref{f6}.  It is seen that both $\delta$ and
$E_0$ decreases with the increase of $\eta_0$.

\section{DISCUSSION}

X-ray and $\gamma$-ray observations of solar flares indicated that
electrons and protons are instantly accelerated in the magnetic
reconnection process. Analyses of HXR spectral data revealed that the
accelerated electrons present a single power-law energy spectrum in most
flares or a double power-law spectrum in some other events.
The co-existence of normal and reverse type III radio bursts
strongly suggests that nonthermal electrons are accelerated near the
reconnection X-point within a compact region less than 2000 km in
length \citep{asch02}. Near the reconnection X-point, both DC electric
field and the reconnection-associated turbulence may contribute to the
acceleration of these nonthermal electrons \citep{stur68,chen07}. The DC
mechanism, as the simplest model, has been studied by various groups
with help of test particle simulations. Although the resulting energy
spectrum of the accelerated electrons is somewhat power law-like, the
spectral indices seem to be much smaller than the typical values derived
from HXR observations. Therefore, more work is required to fill the gap
between test-particle simulations and observations.

Using self-consistent electric and magnetic fields obtained from
numerical simulations of 2.5-dimensional nonlinear MHD equations, we
investigated the energy spectrum of the electrons that are accelerated
by the DC electric field in a reconnecting current sheet. Contrary to
many of the previous studies, our test particle simulations indicate
that the accelerated electrons present a power-law spectrum truncated by
an exponential high energy tail, i.e., $ f(E)\sim E^{-\delta}
e^{-E/E_0}$, which is similar to the case of diffusive shock
acceleration \citep{elli85} and the case of random DC-electric field
acceleration when the electron trapping time is long \citep{anas97}. A
parameter survey is conducted to investigate how various physical
quantities affect the spectral profile and the acceleration rate. As the
energy spectra are fitted with the function $ f(E)\sim E^{-\delta}
e^{-E/E_0}$, it is found that

(1) $E_0$, the rollover energy, increases with larger $B_{g}$ (the
guide magnetic field), larger $L_0$ (which characterizes the size of
the reconnection diffusion region), smaller $\beta_0$ (the plasma 
beta), and smaller $\eta_0$ (the resistivity);

(2) $\delta$ increases with larger $L_0$ and smaller $\eta_0$, and it
saturates as $L_0$ is larger than 100 m or $\eta_0$ is very small.
The dependence of $\delta$ on $B_{g}$ or $\beta_0$ is not monotonic.
$\delta$ reaches the minimum when $B_{g}$ is $\sim 0.4$ times the
antiparallel component of the reconnecting magnetic field, while
$\delta$  reaches the maximum when $\beta_0$ is around 0.01;

(3) The acceleration rate, defined here as the percentage of the
electrons that are accelerated above to 10 keV, increases with larger
$B_{g}$, larger $L_0$, and smaller $\beta_0$. It does not change
much with $\eta_0$, although the percentage of the electrons above
20 keV steadily decreases with increasing $\eta_0$.

The truncated power-law energy spectrum obtained in this paper is
somewhat similar to the double power-law distribution, although the
transition between the low and high energy tails is not so abrupt in
most cases. The research in this paper reveals that, even with
self-consistent electric and magnetic fields, we still cannot reproduce
the observed single or double power-law energy spectra of the nonthermal
electrons in the framework of DC-electric field mechanism. This
discrepancy between test particle simulations and observations might be
reconciled if other effects are included. For example, the turbulence in
the reconnection site would greatly enhance the collision rate between
nonthermal electrons and background particles or waves \citep{wu05}. In
other words, DC-electric field in the reconnection site alone may not be
able to explain the observed spectral features of nonthermal electrons.
The combination of DC-electric field and the turbulence may finally lead
to single or double power-law spectral distributions, which were derived
from HXR observations. This is definitely an issue requiring further
investigation.

\acknowledgements
One of the authors (WJL) thanks V. V. Zharkova and Y. Dai for their
helpful discussions and suggestions. This research is supported by the
Chinese foundations GYHY200706013, FANEDD (200226),
2006CB806302, NSFC (10221001, 10403003, and 10673004).

\clearpage

\clearpage

\end{document}